\documentstyle[12pt, titlepage]{article}

\begin{document}

\title{New Approach to Nonlinear Dynamics of Fullerenes and Fullerites}
\author{G.~M.~Chechin,  O.~A.~Lavrova,  V.~P.~Sakhnenko\\
        (Rostov State University, Russia)\\
        H.~T.~Stokes,  D.~M.~Hatch\\
        (Brigham Young University, USA)}
\date{ Research Institute of Physics of Rostov State University\\
      344090, Stachki Av., 194, Rostov-on-Don, Russia.\\
      chechin@phys.rnd.runnet.ru }

\maketitle

\begin{abstract}
New type of nonlinear (anharmonic) excitations --- {\it bushes} of 
vibrational modes --- in physical systems with point or space symmetry 
are discussed. All infrared active and Raman active bushes for 
$C_{60}$~fulerene are found by means of special group-theoretical methods.
\end{abstract}

 \section{Introduction}
\label{Introduction}

 Vibrations of many fullerenes and fullerites were investigated by 
different experimental and theoretical methods (see~\cite{bib1, 
Rev.paper} and references in these papers). Although the majority of such 
studies are based on the harmonic approximation only, some nonlinear 
(anharmonic) effects also were discussed in a number of papers. For 
example, combination modes of second order brought about by anharmonicity 
of interactions in $C_{60}$~fullerene are discussed and the appropriate 
lines in infrared transmission spectra are reported in~\cite{Martin}.
The above effects does not exhaust the influence of anharmonicity on the 
fullerene and fullerite vibrational spectra, and we want to consider an 
application for these objects of the consistent group-theoretical approach 
for studying nonlinear vibrations in arbitrary physical systems with discrete
 (point or space) symmetry developed in~\cite{Dan1, Dan2, PhysD, 
IntJournNon-LinMech}. 
This approach reveals the existence of new nonlinear dynamical objects 
(or new type of anharmonic excitations) in systems with discrete symmetry 
which we call {\it bushes of normal modes}. The concept of the bush of 
modes can be explained as follows.

In the frame of the harmonic approximation, the set of normal modes can be 
introduced which are classified by irreducible representations (irreps) of 
the symmetry group~$G$ of the considered physical system in equilibrium. In 
this harmonic approximation, normal modes are {\it independent} of each 
other, while interactions between them appear when some anharmonic terms in 
the Hamiltonian are taken into account. Let us note that a very specific 
pattern of atomic displacements corresponds to each normal mode. As a 
consequence, we can ascribe to a given mode a definite symmetry group~$G_D$ 
which is a subgroup of the symmetry group~$G$. The group~$G_D$ is 
a symmetry group of the instantaneous configuration of our system in its 
{\it vibrational} state.

Let us excite at the initial instant $t_0$ only one, arbitrarily chosen mode 
which will be called the {\it root} mode. We suppose that all other modes at 
the initial moment have zero amplitude. Let the symmetry group~$G_D$ and 
irrep~${\mit\Gamma}_0$ correspond to this root mode. Then we can pose the 
following question: to which other modes can this initial excitation 
spread from the root mode? We will refer to these initially 
``sleeping" modes, belonging to the different irreps~${\mit\Gamma}_j ~
(j \ne 0)$, as {\it secondary} modes. 

A very simple answer to the above question was found in~\cite{ Dan1, Dan2}.
It turns out that initial excitation can spread from the root mode only 
to those modes whose symmetry is {\it higher than} or {\it equal to} the 
symmetry group~$G_D$ of the root mode. We call the complete collection 
containing 
the root mode and all secondary modes corresponded to it a {\it bush} 
of modes. Since no other modes are excited, the full energy is {\it trapped} 
in the given bush.
As a consequence of the above idea, we can ascribe the symmetry group~$G_D$ 
(remember that this is a group of the root mode) to the whole bush, and in 
this sense we can consider the bush as a geometrical object.

It was proved in~\cite{Dan1, Dan2, PhysD} that all modes belonging to a given 
bush~B~$[G_D]$ are coupled by {\it force} interactions.
It is very important that the structure of a given bush is {\it independent} 
of the type of interactions between particles of our physical system.

A bush of normal modes can be considered as a {\it dynamical} object, as well. 
Indeed, the set of modes corresponding to a given bush~B~$[G_D]$ does not 
change in time, while the {\it amplitudes} of these modes do change. 
We can write exact dynamical equations for the amplitudes of the modes 
contained in the bush~B~$[G_D]$, if interactions between particles of our 
physical system are known. Thus, the bush~B~$[G_D]$ represents a dynamical 
system whose dimension can be essentially less than that of the original 
physical system.

The above properties of bushes of normal modes can be summarized in the 
following manner. A normal mode represents a specific dynamical regime in a 
{\it linear} physical system which, upon being exciting at the initial 
instant $t_0$, continues to exist for any time~$t > t_0$. Similarly, 
a bush of normal modes represents a specific dynamical regime in a 
{\it nonlinear} system which can exist as a certain object for any 
time~$t > t_0$.

\section{Some mathematical aspects of bushes of normal modes}

Let us examine a nonlinear mechanical system of $N$ mass points (atoms)
whose Hamiltonian is described by a point or space group~$G$. Let
three-dimensional vectors~${\bf x}_i(t)~(i = 1, 2, \dots, N )$
determine the displacement of the $i$-th atom from its equilibrium
position at time~$t$. The $3\times N$-dimensional 
vector~${\bf X}(t) = \{ {\bf x}_1(t), {\bf x}_2(t),\dots,
{\bf x}_N(t) \}$, describing the full set of atomic 
displacements, can be decomposed
into the basis vectors (symmetry-adapted coordinates) of all 
irreps~$\mit\Gamma_j$ of the group G
contained in the {\it mechanical}
representation\footnote{Considering vibrational regimes only, we
can treat~$\mit\Gamma$ as a $3N - 6$ {\it vibrational}
representation of the group~$G$.}~$\mit\Gamma$:
\begin{equation}
\label{eq.21} {\bf X}(t) = \sum_{ji} \mu_{ji}(t)
\mbox{\boldmath$\varphi$}^{(j)}_i = \sum_j
\mbox{\boldmath$\mit\Delta$}_j.
\end{equation}
Here $\mbox{\boldmath$\varphi$}^{(j)}_i$ is the $i$-th basis
vector of the $n_j$-dimensional irrep~$\mit\Gamma_j$. The time
dependence of ${\bf X}(t)$  is contained only in the
coefficients~$\mu_{ji}(t)$ while the basis vectors are time
independent. Thus, a given nonlinear dynamical regime of the
mechanical system described by the concrete vector~${\bf X}(t)$
can be written as a sum of the
contributions~$\mbox{\boldmath$\mit\Delta$}_j$ from the individual
irreps~$\mit\Gamma_j$ of the group~$G$. 

Each vibrational regime ${\bf X}(t)$, can be
associated with a definite subgroup~$G_D$ ($G_D \subseteq G$) which
describes the symmetry of the {\it instantaneous configuration} of
this system. Now the following essential idea is proposed. The
subgroup~$G_D$ is preserved in time; its elements cannot disappear
during time evolution except in the case of spontaneous breaking
of symmetry which we will not consider in the present paper. This
is the direct consequence of the principle of determinism in
classical mechanics.

Introducing operators~$\hat g \in \hat G$ acting on the
$3N$-dimensional vectors~${\bf X}(t)$, which correspond to
elements~$g \in G$ acting on three-dimensional vectors~${\bf
x}_i(t)$, we can write the above condition of preservation of
$G_D$ as a condition of {\it invariance} of the vector~${\bf
X}$(t) under the action of the elements of the group~$G_D$:
\begin{equation}
\label{eq.23} \hat g~{\bf X}(t) = {\bf X}(t),~g \in G_D.
\end{equation}
Combining \mbox{Eqs.\,(\ref{eq.21})} and (\ref{eq.23}) one obtains (for
the details see~\cite{PhysD}) the following invariance conditions
for individual irreps~$\mit\Gamma_j$:
\begin{equation}
\label{eq.25} (\mit\Gamma_j \downarrow G_D) \, {\bf c}_j = {\bf
c}_j.
\end{equation}
Here $\mit\Gamma_j \downarrow G_D$ is {\it the restriction} of the
irrep~$\mit\Gamma_j$ of the group~$G$ to the subgroup~$G_D$, i.e.
the set of matrices of~$\mit\Gamma_j$ which correspond to the
elements~$g \in G_D$ only. The $n_j$-dimensional vector~${\bf
c}_j$ in \mbox{Eq.\,(\ref{eq.25})} is  {\it the invariant vector} in
the carrier space of the irrep~$\mit\Gamma_j$ corresponding to the
given subgroup~$G_D \subset G$. Note that each invariant 
vector of a given irrep~$\mit\Gamma_j$ determines a certain
subspace of the carrier space of this representation, and the
total number of arbitrary constants upon which the vector depends
is equal to the dimension of this subspace.
If in solving  \mbox{Eq.\,(\ref{eq.25})} we find that ${\bf c}_j \ne 0$,
then the irrep~$\mit\Gamma_j$ does contribute to the dynamical
regime~${\bf X} (t)$ with the symmetry group~$G_D$. Moreover, the
invariant vector~${\bf c}_j$ determines the explicit form of the
mode of the irrep~$\mit\Gamma_j$ belonging to the bush of modes
associated with the given nonlinear dynamical regime.

We will illustrate the general statements of bush theory with the 
$C_{60}$~fullerene with buckyball structure and with the icosahedral 
symmetry group~$G = I_h$.
There are 10~irreducible representations of dimensions 1~($A_{g}$,
$A_{u}$), 3~($F_{1g}$, $F_{1u}$, $F_{2g}$, $F_{2u}$), 4~($G_{g}$,
$G_{u}$) and 5~($H_{g}$, $H_{u}$) associated with the
group~$I_h$.  The infrared~(IR)
active modes belong to the irrep~$F_{1u}$, and modes which are
active in Raman (R) experiments belong to irreps~$A_{g}$ or
$H_{g}$. 
We found all bushes of modes for $C_{60}$~fullerene. There are 22 
different bushes for this fullerene.
Let us consider the bush~B7 corresponding to the symmetry group~$G_D =
C_{5v} \subset I_h$. Only four irreps~$A_g,
H_g, F_{1u}$ and $F_{2u}$ contribute to it (the appropriate invariant
vectors are zero for all other irreps
of the icosahedral group~$G = I_h$) \footnote{Since some elements of 
the matrices of
multidimensional irreps of the group~$G = I_h$ are
irrational numbers, we keep only three digits after the decimal
point when we write the invariant vectors.}: 
\begin{equation}
\label{eq.26}
\begin{array}{ll}
\mbox{B7}: [\mbox{symmetry}~C_{5v}]:  & 
 A_g~(a) \: \mbox{--} \: I_h, \:
 H_g~(a, 0.577a, 0, 0.516a, -0.258a) \: \mbox{--} \: D_{5d}\\
 & F_{1u}~(0, 0, a) \: \mbox{--} \: C_{5v} , \:
 F_{2u}~(a, 0.258a, 0.197a) \: \mbox{--} \: C_{5v}.\\
\end{array}
\end{equation}
The arbitrary constants entering into the description of different
invariant vectors are not connected with each other. As all
invariant vectors listed in \mbox{Eq.\,(\ref{eq.26})} are one-parametric
(their arbitrary constants are denoted by the same
symbol~$a$ only for clarity), it is clear that the bush~B7
depends on four arbitrary constants (one constant for each of the
four irreps). The structure~\mbox{Eq.\,(\ref{eq.26})} of the bush~B7 shows
that there exist only four
contributions~$\mbox{\boldmath$\mit\Delta$}_j$ to the appropriate
dynamical regime~${\bf X}(t)$. We denote them\footnote{Hereafter
we write the symbol~$j$ of the irrep~$\mit\Gamma_j$ generating the
contribution~$\mbox{\boldmath$\mit\Delta$}_j$  in square brackets
next to symbol~$\mbox{\boldmath$\mit\Delta$}$.} as
$\mbox{\boldmath$\mit\Delta$}[A_g],
\mbox{\boldmath$\mit\Delta$}[H_g],
\mbox{\boldmath$\mit\Delta$}[F_{1u}],
\mbox{\boldmath$\mit\Delta$}[F_{2u}]$. The invariant vectors
listed in \mbox{Eq.\,(\ref{eq.26})} permit us immediately to write the
explicit form of the dynamical regime~${\bf X}(t)$ corresponding
to the bush~B7 by replacing the arbitrary constants with the four
functions of time~$\mu (t), \nu (t), \gamma (t)$ and $\xi (t)$:
\begin{equation}
\label{eq.27}
\begin{array}{l}
{\bf X} (t) =  \mbox{\boldmath$\mit\Delta$}[A_g] +
\mbox{\boldmath$\mit\Delta$}[H_g] +
\mbox{\boldmath$\mit\Delta$}[F_{1u}] +
\mbox{\boldmath$\mit\Delta$}[F_{2u}] = \\
  \mu (t) \mbox{\boldmath$\varphi$}[A_g] + \nu (t)
\{\mbox{\boldmath$\varphi$}_1[H_g] +
0.577\mbox{\boldmath$\varphi$}_2[H_g] +
0.516\mbox{\boldmath$\varphi$}_4[H_g] -
0.258\mbox{\boldmath$\varphi$}_5[H_g] \} + \\
  \gamma
(t)\mbox{\boldmath$\varphi$}_3[F_{1u}] + \xi (t)
\{\mbox{\boldmath$\varphi$}_1[F_{2u}] +
0.258\mbox{\boldmath$\varphi$}_2[F_{2u}] +
0.197\mbox{\boldmath$\varphi$} 3[F_{2u}]\}.\\
\end{array}
\end{equation}

\mbox{Eq.\,(\ref{eq.27})} is a consequence of the relation of the
group~$G$ and its subgroup~$G_D$ only, and now we must take into
account the concrete structure of our physical system to find
the explicit form of the basis
vectors~$\mbox{\boldmath$\varphi$}^{(j)}_i$ of the irreps entering
into \mbox{Eq.\,(\ref{eq.27})}. They can be obtained by conventional
group-theoretical methods, for example, by the projection
operation method. The basis vectors of the irreps determine the
specific patterns of the displacements of all 60 atoms of the
$C_{60}$~fullerene structure. 

It is important to note that each of the irreps~$A_g, H_g, F_{1u}$
and $F_{2u}$ is contained in the vibrational representation of the
$C_{60}$~fullerene {\it several times}, namely, 2, 8, 4 and 5
times, respectively. (These numbers are equal to the numbers of
fundamental frequencies of normal modes associated with the
considered irreps). As a consequence, we
must treat the time-dependent coefficients in \mbox{Eq.\,(\ref{eq.27})} as
vectors of the appropriate dimensions. Because of this we ascribe
a new index~($k$) to the basis vectors determining the number of
times~($m_j$) the irrep~$\mit\Gamma_j$ enters into the vibrational
representation. Each contribution~$\mbox{\boldmath$\mit\Delta$}_j$
``splits" into $m_j$ copies~$\mbox{\boldmath$\mit\Delta$}_{jk}$,
where $k = 1, 2, \dots, m_j$ and, therefore,
\begin{equation}
\label{eq.28} {\bf X} (t) = \sum_j \mbox{\boldmath$\mit\Delta$}_j
= \sum_j (\sum_{k = 1}^{m_j} \mbox{\boldmath$\mit\Delta$}_{jk}).
\end{equation}
For the case of bush~B7 we have $
\mbox{\boldmath$\mit\Delta$}[A_g]= \mbox{\boldmath$\mit\Delta$}_1[A_g] +
\mbox{\boldmath$\mit\Delta$}_2[A_g],\;
\mbox{\boldmath$\mit\Delta$}[F_{1u}]= \mbox{\boldmath$\mit\Delta$}_1[F_{1u}] +
\mbox{\boldmath$\mit\Delta$}_2[F_{1u}] +
\mbox{\boldmath$\mit\Delta$}_3[F_{1u}] +
\mbox{\boldmath$\mit\Delta$}_4[F_{1u}], \; etc.\\
$ 

The bush~B7 in the $C_{60}$~fullerene structure forms a
19-dimensional dynamical object: its evolution is described by the
dynamical variables listed below as components of the four
vectorial variables~$\mbox{\boldmath$\mu$} (t),
\mbox{\boldmath$\nu$} (t), \mbox{\boldmath$\gamma $} (t)$ and
$\mbox{\boldmath$\xi$} (t)$:   
$\mbox{\boldmath$\mu$}(t) = [\mu_1(t), \mu_2(t) ]$,
$\mbox{\boldmath$\nu$}(t) = [\nu_1(t), \dots, \nu_8(t)]$, 
$\mbox{\boldmath$\gamma$}(t) = [\gamma_1(t), \dots, \gamma_4(t)]$,
$\mbox{\boldmath$\xi$}(t) = [\xi_1(t), \dots, \xi_5(t)]$. 

Thus, although only four of the ten irreps contribute to the  bush~B7,
 its dimension is equal to 19 because several copies of each of these 
four irreps are contained in the full vibrational representation of 
$C_{60}$~fullerene.
We cannot predict the concrete evolution of the amplitudes
of the bush modes without specific information of the nonlinear
interactions in the considered physical systems, but we can assert 
that there does exist an {\it exact} nonlinear regime which involves only
the modes belonging to a given bush.

\section{Optical bushes for $C_{60}$~fullerene}

As was already noted, there are 22 bushes of vibrational modes for 
$C_{60}$~fullerene structure. Five of them are infrared active and 
six are Raman active. We call these bushes by the term ``optical". The 
{\it root} modes of the optical bushes belong to the infrared active 
irrep~$F_{1u}$ or to the Raman active irreps~$A_g$ and $H_g$.
We want to emphasize that some modes associated with the irreps which are 
not active in optics can be contained in a given optical bush.

All optical bushes with their symmetry groups (in square brackets), 
numbers of irreps contributing to them and their dimensions (in parentheses) 
are listed below.

{\sl Infrared active bushes}:

\begin{tabular}{lll}
B7~$[C_{5v}]~(4, \,19)$; & B11~$[C_{3v}]~(6, \,31)$; &B15~$[C_{2v}]~(7, \,46)$; \\   
B19~$[C_s]~(9, \,89)$; & B22~$[C_1]~(10, \,174)$.  &\\   
\end{tabular}

{\sl Raman active bushes}:

\begin{tabular}{lll}
B1~$[I_h]~(1, \,2)$; & B4~$[D_{5d}]~(2, \,10)$; &B5~$[D_{3d}]~(3, \,16)$; \\   
B10~$[D_{2h}]~(3,\, 24)$; & B16~$[C_{2h}]~(5, \,45)$; & B20~$[C_i]~(5, \,87)$. \\   
\end{tabular}

Supposing \rule{0pt}{7mm} that nonlinearity of the considered system is 
{\it weak}\footnote{According to results obtained in the paper~\cite{Martin} 
this hypothesis valid for $C_{60}$~fullerene vibrations.} we can estimate 
the relative values of the contributions from different irreps to a given 
bush. For example, for above discussed infrared active bush~B7 we have
$$
\begin{array}{l}
\mbox{\boldmath$\mit\Delta$}[F_{1u}](root) = O(\varepsilon); \;
\mbox{\boldmath$\mit\Delta$}[F_{1u}](secondary) = O(\varepsilon^3);\\
\mbox{\boldmath$\mit\Delta$}[F_{2u}] = O(\varepsilon^3);  \;
\mbox{\boldmath$\mit\Delta$}[A_g] = O(\varepsilon^2); \;
\mbox{\boldmath$\mit\Delta$}[H_g] = O(\varepsilon^2).\\
\end{array}
$$
Here $\varepsilon$ is an appropriate small parameter characterizing the 
value of the root mode.

Thus, in the case of weak nonlinearity, the contributions of different 
irreps can be of essentially different value. This property seems to be 
important for the interpretation of the vibrational spectra of bushes 
of modes.

\section{Conclusion}

In the present paper, we consider a new type of possible nonlinear 
excitations --- bushes of normal modes --- in vibrational spectra of 
fullerenes and fullerites, using as an example the $C_{60}$~buckyball 
structure.
We believe that special experiments for revealing the bushes of vibrational 
modes in their {\it pure form} will be important for further elucidation of 
the role of these fundamental dynamical objects in various phenomena in 
fullerenes and fullerites. It seems that such experiments may be similar to 
those by Martin and others reported in \cite{Martin}. However, unlike these 
experiments, we must use the monochromatic incident light with frequency 
close to that of the root mode and with polarization along the symmetry axis 
of the chosen bush.

The first-principle calculations are desirable for obtaining the coefficients 
of the anharmonic terms in the $C_{60}$~fullerene for a more detailed 
description of the bush dynamics.

The concept of bushes of normal modes and the appropriate mathematical 
methods for their analysis are valid for both molecular and crystal 
structures. Such a possibility can simplify the assignment of the different 
optical lines in fullerites brought about by both intra- and inter-vibrations
 of the $C_{60}$~molecular clusters.

It will be very interesting to study interactions between bushes of 
vibrational modes and electron subsystems in fullerenes and fullerites.


\begin{thebibliography}{99}
\bibitem{bib1}
C.~H.Choi, M.~Kertesz, L.~Mihaly. $\,$  J.~Phys.~Chem.~A.
{\bf 104}, 102, \{2000\}.
\bibitem{Rev.paper}
H.~Kuzmany, R.~Winkler, T.~Pichler. $\,$ 
J.~Phys.:~Condens.~Matter. {\bf 7}, 6601, (1995).
\bibitem{Martin}
M.~C.~Martin, X.~Du, J.~Kwon, L.~Mihaly. $\,$ 
Phys.~Rev.~B. {\bf 50}, {\it 1}, 173, (1994).
\bibitem{Dan1}
V.~P.~Sakhnenko, G.~M.~Chechin. $\,$ Dokl. Akad. Nauk 
{\bf 330}, 308, (1993). [Phys. Dokl. {\bf 38}, 219, (1993)].
\bibitem{Dan2}
V.~P.~Sakhnenko, G.~M.~Chechin. $\,$ Dokl. Akad. Nauk 
{\bf 338}, 42, (1994). [Phys. Dokl. {\bf 39}, 625, (1994)].
\bibitem{PhysD}
G.~M.~Chechin, V.~P.~Sakhnenko.$\,$ Physica~D 
{\bf 117}, 43, (1998).
\bibitem{IntJournNon-LinMech}
G.~M.~Chechin, V.~P.~Sakhnenko, H.~T.~Stokes, A.~D.~Smith,
D.~M.~Hatch. $\,$ Int.~J.~Non-Linear~Mech. 
{\bf 35}, 497, (2000).
\end{thebibliography}
\end{document}